\documentclass[nohyper,12pt,letterpaper]{JHEP}
\usepackage{epsfig}

\newfont{\frak}{eufm10 scaled 1200}

\newfont{\Bbb}{msbm10 scaled 1200}     
\newcommand{\mathbb}[1]{\mbox{\Bbb #1}}
\DeclareSymbolFont{AMSa}{U}{msa}{m}{n}
\DeclareSymbolFont{AMSb}{U}{msb}{m}{n}
\let\Box\relax
\DeclareMathSymbol{\Box}{\mathord}{AMSa}{"03}

\def \eqn#1#2{\begin{equation}#2\label{#1}\end{equation}}

\title{More Thoughts on the Quantum Theory of Stable de Sitter Space}
\author{T. Banks${}^*$\\
    Department of Physics and Astronomy - NHETC\\
    Rutgers University\\
    Piscataway, NJ 08540\\
    and\\
    Department of Physics, SCIPP\\
    University of California, Santa Cruz, CA 95064\\
E-mail: \email{banks@scipp.ucsc.edu}}

\abstract{I review and update ideas about the quantum theory of de
Sitter space.  New results include a quantum relation between
energy and entropy of states in the causal patch, which is
satisfied by small dS black holes. I also discuss the
preliminaries of a quantum theory in global coordinates, which is
invariant under a q-deformed version of the de Sitter supergroup.
In this context I outline an algebraic derivation of the CSB
scaling relation between Poincare SUSY breaking and the dS radius.
I also review recent work on infra-red divergences in dS/CFT, as
well as the phenomenology of CSB.  I show that a coincidence been
two scales in the phenomenological model is explained by insisting
on the existence of galaxies.  }


\preprint{\hepth{0503066}\\SCIPP-05/01}

\begin{document}




\section{\bf Introduction}

If string theory is to be a theory of the real world, it is crucial
to understand how supersymmetry (SUSY) is broken. Much of the work
on string theory over the past ten years has ignored this problem,
because it is so difficult.  More recently, the advent of flux
compactifications\cite{sethietal} has allowed various authors to
claim a controlled calculation of meta-stable de Sitter (dS)
``states of string theory".   The result is that one is inevitably
led to a multiple vacuum, eternal inflation picture of the universe,
which has been dubbed ``the string landscape".
 I believe that there are many
reasons to be skeptical of these claims and I have documented my
skepticism in a series of papers\cite{isovacetal}.

The purpose of the present paper is not to reiterate these
arguments, but to present an alternative theory of SUSY breaking
and the cosmological constant, which I have been pursuing since
the beginning of the millenium (actually the fall of 1999). It is
currently the only approach to these important problems besides
the landscape.   Most of what I will present here is not novel,
but I will combine various strands of argument that have not
appeared together previously.  I will also present some new
results, which I will describe below.

The method that I have used in analyzing dS space is that of the
phenomenologist.  I take the robust results of quantum field
theory in curved space-time, and treat them as ``experimental
data" which must be reproduced by a correct quantum theory of
gravity in dS space. I begin by outlining the general requirements
that a quantum theory of stable dS space must satisfy.   In
particular, I will review the arguments of Fischler and myself,
that the theory has a finite number of physical states.   I will
then discuss the Hamiltonian, $H$, of a static observer and see
that it must have a number of bizarre properties.   I will discuss
the extent to which this Hamiltonian can be precisely defined.  I
will conclude that the eigenspectrum of this Hamiltonian is
bounded by the dS temperature. Actual masses of particles and
black holes (which are all meta-stable in dS space) are
eigenvalues of a Poincare Hamiltonian $P_0$ which approximately
commutes with $H$ for Poincare energy much less than that of the
maximal Schwarzschild-dS-Nariai black hole. The requirement that
the thermal density matrix for $H$, coincide with a thermal
density matrix for $P_0$, for this range of eigenstates, gives a
relation between the Poincare eigenvalue and the entropy deficit
of the Poincare eigenspace relative to the dS vacuum.  This
relation is satisfied parametrically by small black holes in dS
space. The last result is one of the new features of the present
paper\footnote{This result was known to L. Susskind.}.

The physics of observational interest in dS space is described by
an approximate S-matrix, which is approximately Poincare
invariant.  In static gauge, the dS generators are not connected
simply to the Poincare group. Therefore, I turn to a discussion of
global dS space, in an attempt to make contact with quantum field
theory in curved space-time. Following \cite{nightmare} I will
suggest that the correct quantum formalism provides both a UV and
an IR cut-off for the field theory picture. This leads to the idea
of quantum deformations of the dS group, first proposed by
Rajaraman\cite{arvind}.   The most developed version of this idea
is that of Guijosa and Lowe\cite{guylow}, who introduced the
critical notion of cyclic representations. I propose a Hilbert
space formalism for dS space in which the fundamental variables
are fermionic creation and annihilation operators transforming in
a reducible representation of the quantum group. I propose a
generalization of this to the dS super-group and sketch an
algebraic derivation of the scaling law relating SUSY breaking to
the cosmological constant, which I conjectured in \cite{tbfolly}
and argued for in \cite{sushor}. Some of these algebraic results
are new, but incomplete.   I emphasize in this section that
generators of the dS group are observer dependent.  The generator
corresponding to the "static Hamiltonian" in global coordinates,
does not have the same spectrum and relation to the Poincare
Hamiltonian, as the generator with the same name in static
coordinates.

Coming at the problem from the other end, I will describe a program
for studying the central claim of cosmological SUSY breaking (CSB)
from the point of view of quantum field theory in curved space-time.
The initial reaction of most physicists to this claim is that field
theory gives no indication of the large renormalization of the
classical formula relating the gravitino mass to the c.c. in
spontaneously broken SUGRA.   I believe that this is because no one
has actually studied a gauge invariant definition of the gravitino
mass in dS space.  I argue that dS/CFT gives us such a definition,
to all orders in the usual semi-classical expansion.   dS/CFT is
incompatible with the claim that dS space has a finite number of
states, but this discrepancy is non-perturbative in the c.c.  To all
orders in perturbation theory, the correlators defined by dS/CFT are
invariant under gauge transformations which vanish in the infinite
past and future of dS space.   The formula relating the scaling
property of boundary Green's functions to the bulk mass term, gives
us a gauge invariant definition of mass.   I argue that the
notorious IR behavior of quantum gravity in dS space might give IR
divergent contributions to the boundary dimensions.  I review
preliminary calculations in non-gravitational theories with
minimally coupled massless scalar fields, which exhibit such
divergent mass renormalizations.   If this persists for the
gravitino mass in SUGRA, one would have exhibited a large correction
to classical formulae.   The arguments of \cite{nightmare} then
imply that what we have found is an anomalous dependence of the
gravitino mass on the cosmological constant.   It is not clear that
higher order field theory calculations can give the correct
$\Lambda$ dependence.

I end the paper with a review of a recent attempt to find a low
energy phenomenological lagrangian which implements
CSB\cite{susycosmopheno}. This model involved two new scales in
order to fit the experimental bounds on the standard model.  The
first was the CSB scale, which is determined by $\Lambda$,
according to $m_{3/2} \sim f_0 \Lambda^{1/4} \equiv {F_G \over
m_P}$.  The second was the scale $M_1$ of a strongly coupled gauge
theory with gauge group ${\cal G}$, and is supposed to be
independent of $\Lambda$ for small $\Lambda$. Both $M_1$ and $
\sqrt{F_G}$ were required to be about $1 TeV$. Here I remark that
if dark matter is a ``baryon" of the ${\cal G}$ theory, then these
two scales are tied together by requiring the existence of
galaxies.   The coincidence of scales is the same as the
cosmological coincidence of dark matter and dark energy densities,
and both are explained by insisting that the theory contain
galaxies.

 Taken together, these results
indicate a clear program for studying the theory of stable dS space
more carefully, and suggest that it might lead to phenomenologically
attractive, and predictive results.

\section{\bf The structure of the static hamiltonian}

Quantum field theory in dS space predicts that a time-like
observer\footnote{I use the word observer to denote a large
quantum system, which has many observables whose quantum
fluctuations can be neglected with a certain accuracy.   I will
later argue that in dS space this accuracy cannot be made
infinite, but for purposes of the present section we can ignore
this point.} can only be in causal contact with some of the states
in the QFT Hilbert space. Thus, the local measurements made by
this observer can at most infer an impure density matrix. Gibbons
and Hawking\cite{gh} showed that this density matrix was thermal,
with temperature \eqn{tgh}{T_{dS} = {1\over 2\pi}
\sqrt{{\Lambda\over 3}}/M_P = (2\pi R)^{-1} ,} where $R$ is the
Hubble radius of dS space\footnote{We will work mostly in four
dimensions.   For reasons to be explained below I believe that
this may be the only dimension in which a quantum theory of dS
space makes sense.}.  This result is easily understood once we
announce that field theory in dS space is defined by the analytic
continuation of Euclidean functional integrals on the sphere.

We will have occasion to use different analytic continuations,
which lead to different coordinates on Lorentzian dS.   The
simplest is the static coordinate patch in which we analytically
continue the azimuthal angle $\phi \rightarrow i {\tau \over R}$.
By a simple spatial change of coordinates, the Lorentzian metric
takes on the form \eqn{static}{ds^2 = - d\tau^2 (1- {r^2 \over
R^2}) + {dr^2 \over (1 - {r^2 \over R^2}) }+ r^2 d\Omega^2 .} Here
$d\Omega^2$ is the metric on the unit two sphere. Since the
Euclidean time is periodic with period $2\pi R$, the Lorentzian
Green's functions are thermal, with temperature, $T_{dS} = {1\over
2\pi R}$.

The other continuation we will use is $\theta_1 = {\pi \over 2} +
i {t\over R},$ where $\theta_1$ is one of the polar angles on the
sphere. This gives global coordinates

\eqn{global}{ds^2 = -dt^2 + R^2 cosh^2 (t/R) d\Omega_3^2 ,} where
$\Omega_3$ is the coordinate on a unit three sphere.

\FIGURE[t]{ \epsfig{file=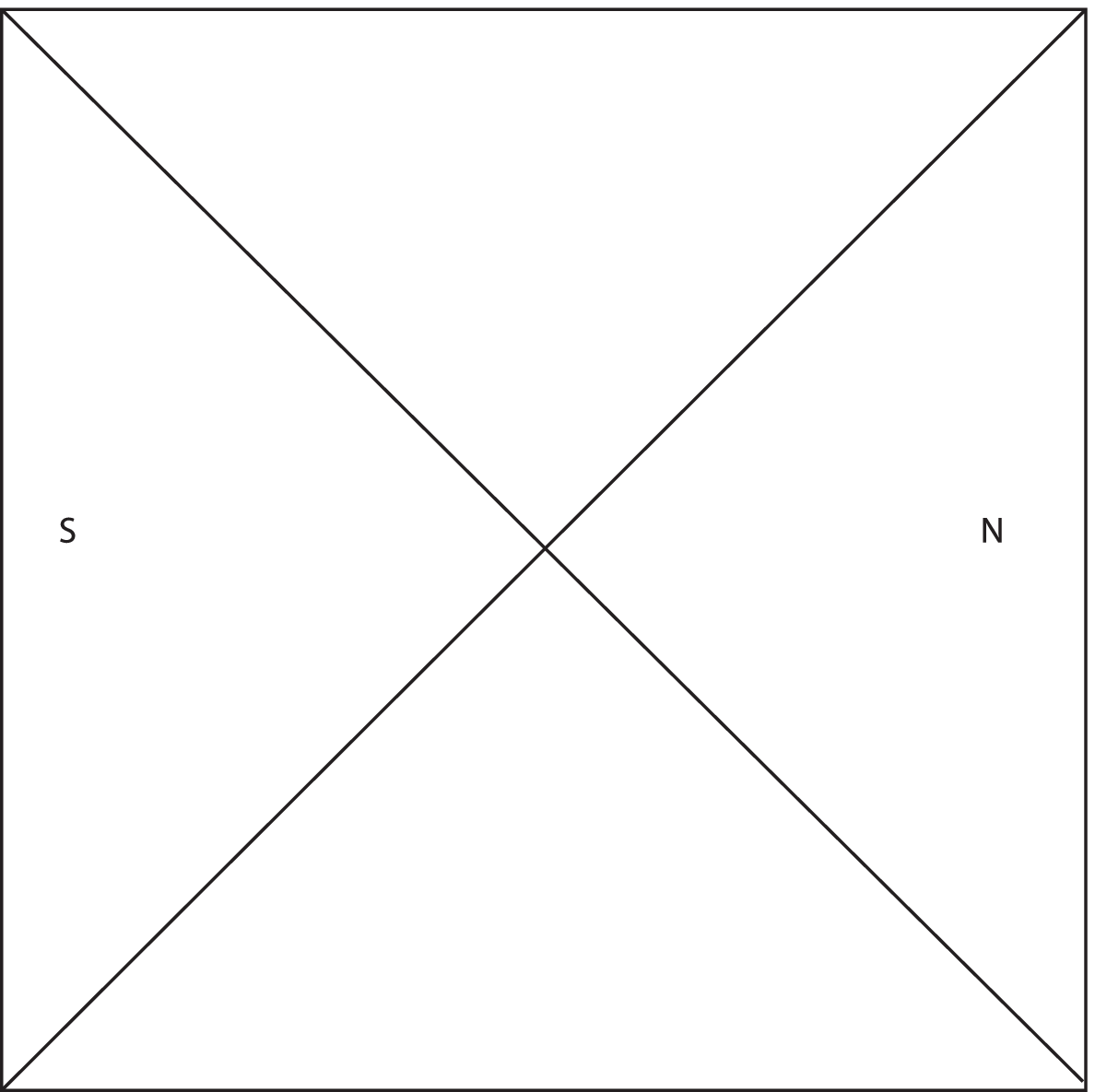,width=0.4\linewidth}
\caption{The Penrose Diagram for de Sitter Space: Time Flows Up in
the North Causal Diamond and Down in the South Diamond.}
\label{dspenrose} }

The Penrose diagram of dS space is the square shown in Figure 1.
The global coordinates cover the whole diagram, while the (North)
static patch covers the triangle labeled $N$. This is the static
patch associated with an observer sitting at a given point (the
North Pole) of the three sphere.   There is an equivalent patch
for any choice of base point\footnote{equivalent in the sense of
being related by dS isometries, which are, in some sense, just
gauge transformations. More on this below.}, in particular, the
South Pole (the cross hatched patch in the picture). At global
time $t = 0$, the entire spatial slice is finite, and it is
covered by the intersection of this slice with the union of the
static patches of North and South poles. Note however that the
directions of static coordinate times in the North and South Poles
are opposite.

There is an interesting quantum mechanical interpretation of these
geometrical facts, which generalizes an observation about black
holes due to Werner Israel\cite{israel}. It has recently been
extended to AdS black holes by Maldacena\cite{maldathermo}, and to
dS space by Goheer {\it et. al.}\cite{goheer}. One can compute
thermal expectation values in a quantum system in terms of
ordinary quantum expectation values in an extended system called
the {\it thermofield double}. One takes two identical
Hamiltonians, $H_+$ and $H_-$ and, in the tensor product Hilbert
space, introduces the Hamiltonian $H = H_+ - H_-$. Now one
introduces the state \eqn{thermodouble}{|\Psi > = \sum e^{-
{\beta\over 2} E_n } |E_n
>_1 \otimes |E_n >_2 . } This state has $H = 0$.  Expectation
values of Heisenberg operators acting only on the first tensor
factor of the Hilbert space, in this state $| \Psi
> $ are equal to time dependent thermal expectation values in the
first tensor factor, with density matrix $\rho = e^{ - \beta H_1}$.

Manifolds like dS space, and black holes, which have analytic
continuations to smooth Euclidean manifolds with compact Euclidean
Killing vectors that fix a point, always seem to have analytic
extensions in which there is a second copy of the region where the
Killing vector is timelike. The second copy has opposite time
orientation. For black holes, Israel\cite{israel} suggested that
quantum field theory on this extended Lorentzian manifold (with
generalized Hartle-Hawking boundary conditions) should be
interpreted as the thermo-field double of quantum gravity in a
static (Schwarzschild-like) patch. This interpretation was extended
to AdS black holes by Maldacena\cite{maldathermo} and to dS space by
Goheer {\it et. al.} \cite{goheer}. Note that with this
interpretation of the physics, the South pole patch of dS space is
no more physical than the extra asymptotically flat universe on the
other side of the Einstein-Rosen bridge in the Kruskal-Schwarzchild
manifold. The physical Hilbert space of the system is just thought
of as the causal diamond of the North Pole.

Another argument which leads to the same conclusion was presented in
\cite{tbfolly}.   In the Euclidean quantization of gravity on
manifolds with the topology of a sphere, rotations of the sphere are
treated as diffeomorphisms and are gauge fixed.   If we want to
analytically continue a Killing symmetry of the sphere and use it as
the Hamiltonian $H$, (the causal patch continuation) we gauge fix in
such a way that $H$ is left as a symmetry of the quantum Hilbert
space.   The rest of the generators are treated as diffeomorphisms.
We mod out by them and they do not act on the physical Hilbert space
({\it cf.} light cone gauge quantization of the string).   On the
other hand, \cite{nappi} in the standard treatment of ordinary
quantum field theory in dS space, by Euclidean methods, we can
analytically continue to the causal patch and then obtain Green's
functions on the global dS manifold by acting with dS
transformations which map the causal patch onto the rest of the
manifold.   This suggests that we view the region outside of a given
causal patch as simply a gauge copy of the region inside of it.
This is one origin of the idea of {\it observer
complementarity}\cite{mcosmoverl}, to which we will return below.

Now imagine, as everyone does, that in the quantum theory of
gravity, there is some kind of UV cutoff of spatial scales. In a
theory with finite volume, we would then expect to have a finite
number of states.  A much stronger argument that the quantum theory
of dS space has a finite number of physical states, comes from the
observation of Gibbons and Hawking that it appears to be a thermal
system, with a finite entropy.   We need one extra assumption to
prove that the number of states is finite, namely that the energy
spectrum of the static Hamiltonian is bounded from above.    Here we
will give only a naive argument for this, without being very precise
about how energy is defined.  Below we will argue that it is very
likely that the upper bound on the spectrum of the static
Hamiltonian is much more stringent than the one we are about to use.

In the QFT approximation, when we look at localized states of very
high energy, we inevitably come to a point that the Schwarzchild
radius of the state is larger than the radius of the state in the
background geometry.   At this point the QFT approximation breaks
down and states are best described as black holes.   The scaling of
entropy with size changes.  The maximal energy is now completely
determined by the size of the state.   Most states of a fixed
energy, with the minimal size for that energy, are black holes.

In dS space there is a maximal black hole, the Nariai hole, whose
cosmological and black hole horizon areas coincide.   The metric of
the causal patch has the form \eqn{caualpatch}{ds^2 = - \beta^2 (r)
dt^2 + {dr^2 \over \beta^2 (r)} + r^2 d\Omega^2} where $\beta^2 (r)
= (1 - r^2/R^2)$.   Black holes in the causal patch are described by
the replacement $\beta^2 \rightarrow \beta_M^2 \equiv (1 - {2M\over
r} - {r^2 \over R^2 })$.   $\beta_M$ has two zeroes, $r_{\pm}$, for
$M>0$, which locate the positions of the black hole ($r_-$) and
cosmological ($r_+$) horizons. Semiclassically, both horizons are
sources of thermal radiation, and the black hole horizon is hotter,
except in the limiting case (the Nariai black hole) in which the two
horizons coincide.   This leads to the expectation that black holes
decay, which is certainly correct for $M \ll R$. Bousso and
Hawking\cite{raphhawk} have argued that the same is true for all
values of $M$ including the Nariai case. This point is probably
worth revisiting, since it will be crucial in the discussion below.

It is also worth noting that the entropy of all of these black hole
states is smaller than that of the empty dS vacuum.   The total area
of the two horizons is smaller than the area of the empty dS
horizon, and monotonically decreases as the mass is increased. This
again suggests that the black holes should be viewed as low entropy
excitations of the dS vacuum, which decay back to it.  The behavior
of the entropy as a function of black hole mass frustrates an
observer's attempt to access all of the states of the system as
localized excitations in its horizon volume.   As the local entropy
is increased, the total entropy gets smaller indicating that an
observation of a large amount of localized entropy freezes the rest
of dS space into a very low entropy state\footnote{This conclusion
is supported by the extended Penrose diagram of dS black holes.  The
global geometry has a black hole in both North and South Pole causal
patches, consistent with the Israel interpretation as a thermofield
double . In the Nariai limit, every observer sees a (generically
moving) black hole in its causal patch.} .    The existence of a
maximal energy in dS space, coupled with the assumed thermality of
the density matrix, and finite entropy, imply that the system has a
finite number of states.

Our observations about the decay of black holes, have important
implications for the structure of the static Hamiltonian.   One
might have thought that the black hole mass parameters referred to
approximate eigenvalues of the dS Hamiltonian.  That is, one
imagines an approximate Hamiltonian,$H_0$, for which black holes are
stable excitations with eigenvalue equal to their mass parameter.
Hawking decay would then be the result of a correction to this
Hamiltonian, but the diagonal matrix element of the Hamiltonian in
the meta-stable black hole state would be approximately equal to the
mass. Indeed, Gomberoff and Teitelboim\cite{gt} have described a
semiclassical formalism in which one can define a conserved
generator in dS-black hole space-times, whose value is the black
hole mass.   I have advocated this as evidence for the above
Hamiltonian interpretation in \cite{davis} . However, if all black
holes decay, this is inconsistent with quantum mechanics\footnote{I
would like to thank L. Susskind, and particularly M.Srednicki, for
arguments, which set me straight on this point.}. In such a picture,
$M$ would be approximately the diagonal matrix element of the true
Hamiltonian in the $M$ eigenstate of $H_0$. However, the diagonal
matrix element of the Hamiltonian is always bounded by its maximal
eigenvalue.  If all black holes decay, the maximal eigenvalue of $H$
is of order the dS temperature.   We conclude that the entire
eigenspectrum of $H$ consists of states in the dS vacuum
ensemble\footnote{It should be noted that if the Nariai black hole
were stable, this conclusion would not follow.}.

Classically, these vacuum states have zero energy, but quantum
mechanically we might imagine that they have energies as high as the
dS temperature.   Indeed, a random spectrum of states, spread
between $0$ and $T_{dS}$ with density $e^{- S_{GH}} = e^{ - \pi
R^2}$ would provide a heat bath for localized states in dS space,
thus explaining why they have a thermal density matrix.  The
Coleman-DeLucia tunneling amplitudes between two different dS
spaces\cite{CDL}, provide further evidence for this identification.
Probabilities for the forward and reverse tunneling processes are
computed in terms of the same instanton solution, but with a
subtraction of the action of the initial dS space for the transition
$dS_I \rightarrow dS_F$. The ratio of probabilities for inverse
processes is given by the exponential of the difference of the
entropies. This is what we expect from the law of detailed balance
if the entropy is the dominant term in the free energy. The
condition for entropy dominance is that most of the states lie below
the temperature.

Similar arguments can be made about the decay of all states of the
system which are not black holes but have energies above the dS
temperature. There are no arguments that such states are stable.
Consider for example an electron in our own universe, assuming that
it asymptotes to a dS universe with the nominal value of the
cosmological constant indicated by cosmological data. We usually
argue that electrons are stable because of charge conservation, but
in dS space, the charge of an electron in the static patch is
canceled by a tiny charge density uniformly spread over the
cosmological horizon.  From the point of view of global coordinates,
this is a consequence of the fact that spatial sections are compact
so the global manifold cannot have a net charge. As a consequence,
electrons in a causal patch of dS space are not stable, even if they
are bound to the observer, because there is a finite probability for
the charge density on the horizon to coalesce as a positron and
annihilate the electron. This probability is extremely small, but
the electron's lifetime is much shorter than the dS recurrence time,
which we will discuss later.

The same argument we used above shows that the electron mass cannot
be an approximate eigenvalue of the static dS Hamiltonian. This
seems paradoxical, and contradicts the predictions of QFT in a dS
background.   There are two elements to what I think is the correct
resolution of this paradox.   The first is fairly conventional.   It
is well known that the conventional constraint equations of GR in a
temporal gauge

\eqn{constraints}{{\cal H} = {\cal P}_i = 0,} can be interpreted
as the vanishing of the total energy and momentum densities once
gravitational effects are taken into account.  Total energy and
momentum are pure surface terms. In the causal patch the only
surface on which to imagine defining these objects is a stretched
horizon, a time-like surface within a few Planck distances of the
horizon.   The QFT description of physics breaks down on such
surfaces, because of the large blue-shift between the observer and
the stretched horizon.   Thus, we can easily imagine, in the full
theory of quantum gravity, near horizon contributions to the total
static energy which almost cancel the QFT contribution. The
electron mass we measure might be viewed as an integral over
surfaces far enough away from the horizon that these near horizon
quantum gravitational corrections are small.

These remarks are a tentative answer to the question of why the
electron energy is ``really" almost zero, but they do not explain
how to define the electron mass we measure in dS space as a
mathematical quantity in a holographic formulation of the theory. I
think that the answer to this is related to the relation between the
approximate Poincare generators, and dS generators of a dS space
with very small c.c.

The near (future) horizon geometry of the causal patch is
\eqn{causnearhor}{ds^2 = R^2 (- du dv + d\Omega^2)} while that of
asymptotically flat space near future null infinity is
\eqn{minknearhor}{ds^2 = - {1\over v^2}(dudv + d\Omega^2).} In both
cases the horizon is at $v \rightarrow 0$.   The static dS
Hamiltonian is the boost $u\partial_u - v\partial_v $ in $(u,v)$
coordinates. Lorentz transformations are conformal transformations
of the near null infinity geometry, while the translation generators
have the form $P_{\mu} = Y_{\mu} (\Omega ) \partial_u$, where $Y_0 =
1$ and $Y_i = n_i$, (with $n_i^2 = 1$ parametrizing the
sphere\footnote{We have chosen a particular Lorentz frame in which
the metric on the sphere at infinity is round.}).

Now imagine that we have a quantum realization of dS space which,
for large $R$ carries both a representation of the static
Hamiltonian and an approximate representation of the Poincare
algebra.   Given our discussion above the static generator should
have a spectral cutoff of order (Planck units) $T_{dS} \sim {1\over
R}$. The Poincare generator, $P_0$ will also be bounded, because the
system has a finite number of states.   Given that the spectrum
above the Planck scale is dominated by black holes, we might imagine
the cutoff on $P_0$ is of order $R^{d - 3}$, the maximal black hole
mass in dS space.    A cartoon version of these generators would be
 \eqn{cartoongen}{H \sim {1\over
R}(u\partial_u - v\partial_v)\ \ \ \ P_0 \sim R
\partial_u ,} with order one cutoffs on the spectrum of the
derivative and scaling operators. Then

\eqn{cartooncom}{[H, P_0] \sim {1\over R} P_0} Thus, for large $R$,
the commutator is small for states whose $P_0$ distribution contains
only eigenstates with energies $\ll R $. For states satisfying this
constraint, $P_0$ becomes an additional approximate quantum number,
whose spectrum breaks the huge degeneracy of $H$. This equation
could also explain why $P_0$ eigenstates are approximately stable
under the time evolution defined by $H$. Masses of meta-stable
particle and black hole excitations of the system would be related
to $P_0$ eigenvalues.

Note that it is only in four dimensions that the $P_0$ generator
approximately commutes with the static generator for energies all
the way up to the maximum black hole mass.  For $d > 4$ the
generators fail to commute for parametrically smaller values of
$P_0$.  The stability of large black holes could not be explained
by approximate conservation of $P_0$.   This might be an
indication that a quantum theory of dS space may only make sense
in four dimensions, or that the commutation relation is modified
in higher dimensions.

This picture implies an interesting relation for the expansion
coefficients of $P_0$ eigenstates in the basis with $H$ diagonal:

\eqn{p0}{\sum | p_0 >< p_0 | = \sum P_n (p_0) | h_n >< h_n | :   \ \
\ H| h_n
> = h_n | h_n >.}  The sum on the left hand side of this equation could be either
the degeneracy of the eigenvalue $p_0$, or the density matrix of
states within an interval of size $T_{dS}$, centered on $p_0$. The
density matrix is $e^{- {H\over T_{dS}}}$. QFT tells us that for
eigenstates of $P_0$ for which a QFT description is valid (including
at least black holes whose radii grow more slowly than $R$ as
$R\rightarrow\infty$) the probability of finding $| p_0 >$ is
approximately $e^{-{{p_0}\over T_{dS}}}$. Thus

\eqn{p0ceffs}{ {\sum P_n (p_0) e^{-{{h_n}\over T_{dS}} } \over
{\sum e^{-{{h_n}\over T_{dS}}}}}
 = e^{-{{p_0}\over T_{dS}}}.} An easy way
to get this result is to write the Hilbert space as a tensor
product.  Thus, we break the index $n$ up into a pair of indices
$n = (k,L)$.   Assume that $\sum_k = e^{{{p_0}\over T_{dS}}}$ and
that $P_{k,L} = \delta_{k1} $.  We will also approximate
$e^{-{{h_n} \over T_{dS}}} \approx 1$.   Thus we get a relation
between the $p_0$ eigenvalue and the entropy deficit of this
eigen-space relative to the whole Hilbert space.

An exactly similar relation appears for black hole entropy for
masses $\ll R^{d-3}$.   The cosmological horizon for such black
holes occurs in static coordinates at

\eqn{cosmhor}{r_+ \approx R -{8\pi M\over{(d-2) A_{d-2}R^{d-4}}}.}
The area deficit, relative to empty dS space is
\eqn{deltaA}{\Delta A = (d - 2) A_{d-2} MR} where $A_{d-2}$ is the
area of the unit $d - 2$ sphere.   The relation between entropy
and mass given by this equation, coincides with the one derived
quantum mechanically in the previous paragraph\cite{raphbek}.
Interestingly, the same argument works for charged black holes.
The entropy formula depends on the charge, but the leading
correction for $R_S \ll R$ does not.  Note that this is a
requirement for agreement with the thermal formula from quantum
mechanics, which makes no reference to the charge.

In \cite{davis} I reported on a formulation of dS quantum
mechanics in terms of fermion operators (see the next section) and
proposed to use this definition of mass in terms of entropy to
define the static Hamiltonian.  This is now seen to be wrong, but
it may be possible to consider that construction to be a model of
the black hole eigenstates of the Poincare Hamiltonian .

The basic idea of that model, which was constructed in collaboration
with B. Fiol, was simple: introduce fermion operators
$$[\psi_i^A , (\psi^{\dagger})_B^j ]_+ = \delta_i^j \delta_B^A$$
which are $N \times N + 1$ matrices.   These are ``operator valued
sections of the spinor bundle" over the holographic screen on the
cosmological horizon of a given observer\cite{susyholoscreen} and
represent quantized pixels on the screen.  As we will see in the
next section, $N$ is proportional to the dS radius.

 We work in the
approximation in which the static Hamiltonian is the unit matrix,
and will not reproduce the commutator of static and Poincare
generators or the instabilities of black holes.   The idea is to
implement the principle of {\it asymptotic darkness}: start from an
approximation to quantum gravity in which the simple states are
stable black holes, which dominate the high energy spectrum, and
introduce black hole instabilities as a perturbation. We want to
describe the basis of black hole eigenstates of the Poincare
Hamiltonian.

The entire Hilbert space is the dS vacuum ensemble of the static
Hamiltonian.   We identify black hole states by choosing an integer
$N_- < [{N\over 2}]$.   Now, in some specific basis for the
fermionic matrices, make a block decomposition
$$\pmatrix{\Psi_+ & \Psi_{+-} \cr \Psi_{1+} &\Psi_{1-} \cr \Psi_{-+} & \Psi_-}.$$
$\Psi_{\pm}$ is an $N_{\pm}\times (N_{\pm} + 1)$ matrix, with $N_+
= N - N_- - 1$.  $\Psi_{ 1\pm}$ is a $1 \times N_{\pm}$ matrix,
and $\Psi_{-+}$ and $\Psi_{+-}$ are $N_- \times N_+ + 1$ and $N_+
\times N_- +1$ matrices respectively.   Black hole states are
identified by the constraint \eqn{bhconst}{(\Psi_{-+})_i^A | BH >
= \Psi_{1-} | BH > = 0.} The states created by $\Psi_+$ and
$\Psi_{+-}$ creation operators should be identified with the
cosmological horizon in the presence of the black hole, while
those created by $(\Psi_-)^{\dagger} $ operators are associated
with the black hole horizon.   The states created by
$(\Psi_{-+})^{\dagger} $ operators should perhaps be associated
with particle states propagating in the space between the two
horizons, but we will see that this does not account for the
entropy of such states.   The $ 1 \times N_{\pm}$ matrix fermions
do not have any obvious macroscopic interpretation.

The equations for the horizons of a Schwarzchild dS black hole may
be put in the form
$$R^2 = (R_+ + R_-)^2 - R_+ R_- \approx R_+2 + R_+ R_-$$
$$ 2M R^2 = R_+ R_- (R_+ + R_-) \approx R^2 R_- , $$ where the
approximate forms are good in the limit $R_+ \gg R_-$.   The entropy
deficit of a small black hole is $\pi R R_- = 2\pi R M$.   In the
fermion model the entropy deficit is approximately $N N_-$ for $N
\gg N_-$, so if we identify (in Planck units) $\sqrt{\pi} R = N$ and
$\sqrt{\pi} R_{\pm}  = N_{\pm}$, our model reproduces the ``data".
We also obtain the identification $M = {1\over 2\sqrt{\pi}} N_-$.

In order to write an operator form for the Poincare Hamiltonian,
we must choose a matrix basis for the maximal size black hole $N_-
= [{N\over 2}]$ .   Black holes of a given size are constructed by
imposing the above constraint for a given value of $N_-$ less than
the maximum.   The Poincare Hamiltonian is given by
$$M = {1\over 2\sqrt{\pi}} ([{N\over 2}] - \sum_{j,A} [(\Psi_{-+})^{\dagger}]_A^j
[\Psi_{-+}]_j^A) ,$$  where $1 \leq j \leq [{N \over 2}]$ and $1
\leq A \leq N - [{N \over 2}] - 1$.  Note that this formula does
correctly reproduce the fact that what we have called particle
excitations of a small black hole make positive contributions to
the energy, but it does not reproduce their spectrum or entropy.
I view this as crudely analogous to the way that the free quark
gluon Hamiltonian correctly describes generic high energy
excitations of QCD, but fails to reproduce the hadron spectrum.

An important feature of the above construction is that our
description of the black hole spectrum involved a choice of basis
for the fermionic matrices.   Other choices will give the same
results, but the states that one ``observer" associates with a
localized black hole, will be mixed up with the cosmological horizon
states of an ``observer" who makes a different choice of basis. It
is very tempting to identify this ambiguity as the quantum analog of
the choice of static coordinate frame in classical de Sitter space.

To summarize, it appears that the spectrum of the static dS
Hamiltonian is cut off at an energy of order $1/R$, with a density
of states of order $e^{-S}$ where $S$ is the dS entropy. $e^{S}$ is
not exactly the number of states, because the density matrix is
thermal. However, since most of the states are at energies below
$T_{dS}$, the discrepancy between $e^S$ and the number of states is
not large, and goes to zero as $R \rightarrow\infty$. The huge
degeneracy of dS horizon states is broken by another operator,
$P_0$, whose commutator with the static Hamiltonian is small in the
subspace spanned by eigenvectors of $P_0$ with eigenvalue $\ll R$.
$P_0$ is one of the generators of a super-Poincare algebra, which
emerges in the limit $R \rightarrow\infty$.  The requirement that
the thermal density matrix of the static Hamiltonian also gives
thermal statistics for the Poincare generator, implies a relation
between the Poincare eigenvalue $p_0$, and the entropy deficit
relative to the dS vacuum of the $p_0$ eigenspace.   This relation
coincides with that derived from the Bekenstein-Gibbons-Hawking
formula for small dS black holes.  We constructed an explicit
quantum model, which incorporated this relation.

An important consequence of this discussion, is that the
information, which is of concern to particle physicists, is
encoded in the Poincare Hamiltonian, rather than the static dS
Hamiltonian.  The two are related by the commutation relation $[H,
P^0 ] \sim {1\over R} P^0 $.  As we will see, the part of the
$P^0$ spectrum where $P^0$ is an approximately conserved quantum
number under $H$ evolution, accounts for only a tiny fraction of
the states in the Hilbert space.  The rest of the states are not
localizable in the observer's horizon, and form a degenerate soup
which is localized close to the horizon. From the static
observer's point of view\footnote{Which means from the point of
view of any realistic measurements.} the description of the rest
of the Hilbert space is somewhat arbitrary.   It must satisfy some
weak constraints, which ensure that the states on the horizon
thermalize the localizable degrees of freedom at the right
temperature.  In later sections we will see how to obtain a more
constrained description of dS space by choosing a basis for the
Hilbert space which simultaneously describes localizable states in
disjoint, causally disconnected horizon volumes.

\subsection{Measurement theory in dS space}

Working physicists usually try to avoid thinking about the arcana of
quantum measurement theory.   Discussions of these issues often
smack of academic philosophy, and we all know what we really do to
measure something anyway, right?

I'm afraid that in the quantum theory of gravity we really have to
address these issues.  What is more, discussion of them illuminates
certain otherwise obscure features of the theories that we know to
make mathematical sense, namely the fact that the only gauge
invariant observables are boundary correlators. Finally, I will
argue below that a question which properly belongs to the philosophy
of science has some relevance for decisions about whether a quantum
theory of dS space can ever make sense.

Anyone who has ever thought about quantizing generally covariant
field theories knows that there is a problem of interpreting the
observables in terms of local physics on the world volume of the
universe.   In $0+1$ and $1+1$ dimensions, where Wheeler-DeWitt
quantization makes sense beyond the semiclassical approximation, the
observables refer to global properties of multiple disconnected
``universes" (world lines or world sheets).  They do have an
interpretation as the perturbation expansion of a Scattering matrix
(and in the particle case the local correlation functions) of an
external space-time in which the particles and strings are embedded,
but no interpretation in terms of approximately local physics on the
world volume.

Hawking suggested many years ago\cite{hawk} that the S-matrix would
be the only gauge invariant observable in a theory of quantum
gravity in asymptotically flat space-time\footnote{And then
proceeded to argue that the S-matrix would be replaced by the Dollar
Matrix.}. Modern developments, particularly the Fischler Susskind
Bousso covariant entropy bound suggest a deep reason for this, which
is connected to quantum measurement theory.

Quantum measurement theory can be summarized in a few lines in the
following way:  Certain large quantum systems, in particular cut-off
quantum field theories, have {\it pointer observables} whose quantum
fluctuations can be made very small.  The canonical example is the
volume averaged value of a local field. There are states of the
system in which the quantum fluctuations of such pointer observables
are arbitrarily small in the limit of infinite volume.  The system
has a large number of states with essentially the same value for the
pointer observable. Typical tunneling amplitudes from one value of
such a {\it pointer observable} to another, are of order $e^{ -
VM^3}$ where $M$ is the energy cutoff and $V$ the volume over which
the observable is averaged.

Quantum mechanics makes precise mathematical predictions for any
system.   The operational meaning of these precise predictions is
extracted by coupling the original system to a measuring apparatus
in such a way that the value of a microscopic observable $A$ is
correlated with the value of a pointer observable of the
apparatus.   These measurements are robust over time scales of
order the inverse pointer tunneling probability per unit
time\footnote{ as long as we do not make another measurement on
the same system, which measures a variable $B$ that does not
commute with $A$.}. Infinite precision and robustness are attained
by taking the limit of an infinite measuring apparatus.

These ideas are problematic in a theory of gravity, because
infinite machines have infinite gravitational interaction with the
measured system.  The only way to resolve this problem is to make
the measurements at infinity, and this is the reason that string
theory, the only quantum theory of gravity which specifies a
complete set of gauge invariant observables, only makes
predictions about scattering amplitudes and other kinds of
boundary correlators, in space-times which are globally foliated
by infinite volume spatial sections.

In a theory of a stable dS universe, we do not have the luxury of an
infinite boundary of space on which to make measurements.   Indeed,
I have argued here that we can describe an entire stable dS universe
with a finite number of physical states.   {\it This means that the
theory of such a space-time has an inherent quantum uncertainty
built in to it.}   The theory cannot self consistently describe
measurements of its predictions with a greater accuracy and
robustness than some fixed finite bound.  This means, that there
cannot be a unique mathematical description of the theory.   Two
mathematical models, whose predictions differ by an amount smaller
than this {\it a priori} bound on the precision of measurements,
will not be operationally distinguishable. Predictions over time
scales greater than the tunneling time for the most robust pointer
observable that can be manufactured from the ingredients at hand,
are meaningless.

There is a subtle point of scientific philosophy inherent in the
statements of the previous paragraph.  It has to do with the precise
relation between mathematics and the physical world.   The
mathematical elegance of the known laws of physics, has lead many
researchers to the (at least subconscious) conclusion that the
relation is Platonic/Pythagorean.   That is, mathematics has some
kind of existence outside the world of measurement, so that the
predictions of a mathematical theory should be discussed and taken
seriously, even when we cannot, in principle, devise a method for
testing them.

My own point of view is quite different.  Mathematics is a creation
of human beings, who are physical objects in a world we can know
about only through measurement. We can only discuss those of its
predictions which are, at least in principal, subject to
verification by physical observation. Mathematics produces models.
If they are models of the entire universe then they must supply us
with a self consistent set of instructions for testing the model.
Predictions which go beyond the model's ability to ``self-test" have
no physical meaning.

In \cite{nightmare} we proposed that this should be viewed as a kind
of gauge invariance: two Hamiltonian descriptions of {\it e.g.} the
static coordinate patch of dS space are equivalent if their
predictions for all observables agree within the intrinsic
limitations on the accuracy of measurements\footnote{It is becoming
more and more tempting to imagine that this {\it quantum measurement
gauge equivalence} lies at the root of the general covariance of
classical GR.  That is, we will eventually derive the latter
principle from the former.}.

With the philosophical baggage out of the way, we can try to discuss
what the quantitative bounds on precision in dS space might be.
There are three qualitatively different types of states in the dS
Hilbert space, when it is viewed from the perspective of the static
patch observer.  The first class of states are well described by
quantum field theory in the static patch.  To get a qualitative idea
of their number one notes that most of the states in a 4 dimensional
QFT are described by the conformal fixed point theory.   We must
cut-off the field theory at some scale $M$ and put it in a box of
radius $\sim R$, so the entropy of field theoretic states is of
order $M^3 R^3$.   The energy of a typical state of this type, will
be of order $M^4 R^3$, and this is also the Schwarzschild radius of
the state in Planck units.   Insisting that $R > R_{Schw}$ we find
$M < R^{-{1\over 2}}$, which implies that the entropy in field
theoretic states is $ < R^{3/2}$ in Planck units.

The second class of states are the states on the cosmological
horizon.   In fact, as we have discussed above, these are the only
absolutely stable eigenstates of the static patch Hamiltonian. Their
entropy is of order $R^2$. These states are not useful for making a
measuring apparatus for the static observer. A static observer can
only access them by probing a region of his coordinate patch with
very large quantum/thermal fluctuations.

Finally, we have horizon states of black holes localized within a
causal patch. Their entropy is also of order $R^2$ (for black
holes of order the horizon size), though with a coefficient one
third of the dS vacuum entropy. The No-Hair theorem suggests that
it is not possible to build robust pointer observables from black
hole eigenstates. Although they are numerous, they are much more
degenerate than states of a quantum field theory, and the locality
arguments, which guarantee small tunneling amplitudes between
different macrostates of a QFT, do not apply to them.   Of course,
our understanding of black hole microstates is still too primitive
to make definite conclusions about this point.

Assuming that robust measuring devices can be constructed only
from QFT states, we can draw two important conclusions about the
description of physics in the static patch of dS space.   The
first is that we cannot access most of the states of the theory by
measurements.   QFT can only tell us inclusive things about the
horizon states.   The semi-classical analysis that we have
reviewed above suggests that, as far as QFT states are concerned,
the rest of dS space imitates a thermal bath at the dS
temperature.   Thus, models of the static Hamiltonian which differ
only in their description of the horizon states will be
essentially equivalent if they couple the horizon states to the
QFT states in a way that is compatible with thermalization. The
second point is that the tunneling time for the most robust QFT
devices is, for large $R$, an infinitesimal fraction of the dS
recurrence time\cite{recur}. Indeed, for the value of $R$
indicated by cosmological observations in the real world, the
recurrence time is essentially the same number in Planck units
($e^{10^{123}}$) that it is in units of this tunneling time
$(e^{10^{123} - 10^{92}})$ .  These two remarks are connected.
The recurrence time is related to the splitting between the levels
of the horizon states. Many different Hamiltonians for the horizon
states will give the same results for physics measured by QFT
devices, but predict different horizon configurations over a
recurrence time. It is only by paying attention to these
unobservable properties of the model that we can attribute reality
to the recurrence time scale.

The considerations of this section lead to another interesting
observation.   The total number of degrees of freedom, which can be
described by local field theory in a fixed horizon volume, is much
less than the total entropy of dS space.   This means that the dS
Hilbert space contains enough states to describe of order $R^{1/2}$
horizon volumes with independent (commuting) degrees of freedom. In
the global picture of dS space in QFT, the system at late global
time seems to have an infinite number of copies of the degrees of
freedom of a single horizon.   These considerations suggest an IR
cutoff on the QFT Hilbert space.   Crudely speaking, this is an
upper cutoff on the value of global time, of the form $(
cosh(t_{max} /R))^3  \sim (R M_P)^{1/2}$.  Of course, if we are
willing to reduce the UV cutoff on our field theoretic states, we
can use a QFT description of most of the states at larger values of
the global time\footnote{We can also use QFT with a relatively large
cutoff, but insist that the system is in its vacuum over most of
space-time.  All of these descriptions should eventually be thought
of as different gauge choices.}.

In the next section, I will try to give a more precise definition of
a finite system which might converge, as $R\rightarrow\infty$, to
the QFT picture of dS space in global coordinates, at a global time
where there are of order $(RM_P)^{1/2}$ horizon volumes in the
spatial slice.

\section{Global coordinates and local physics}

We have seen that a local observer in dS space can only access a
limited fragment of the information content of the space-time.
This implies that there is large equivalence class of choices for
the Hamiltonian of the static observer, which will give rise to
the same predictions for experiments we are likely to do. It is
reasonable to search for members of this equivalence class, which
are simple and mathematically elegant, and make choices about how
to describe those aspects of the physics a local observer will not
access.

The ambiguity in the quantum description of dS space is related to
the question of what the $SO(1,4)$ isometry group of dS space means.
Witten\cite{witstrom} has argued that it is just a group of gauge
transformations, which does not act on gauge invariant physical
states.  Our arguments suggest that there are no precise
measurements in dS space, and that the most precise measurements one
can contemplate are tied to the frame of a particular local
observer.   Thus, it seems clear that the $ R \times SO(3)$
generators that preserve a given causal patch, are to be viewed as
global symmetry generators.   We have seen however that much of the
interesting local physics is encoded into the Poincare Hamiltonian,
which is only an approximate symmetry, rather than the static
Hamiltonian.  A more global description of dS space leads to the
possibility of a more elegant description of the physics, in which
the Poincare Hamiltonian is more closely tied to the dS group.

In a global description of dS physics, the arguments of
\cite{nightmare} and of the previous section, suggest that there
will be something resembling a field theoretic description of all
states. In field theory dS isometries converge to the Poincare
group, and the distinction between the generators that we
introduced in the static gauge is no longer apparent.  Recall that
the description of the relation between static and Poincare
Hamiltonians was motivated by the behavior of these generators on
the horizon.  A global description should have no trace of a given
observer's cosmological horizon. It depicts most of the states of
the system as (cutoff) localizable field theory states on the dS
manifold at a time sufficiently later than $t=0$, that they do not
form black holes. The two descriptions can only agree
(approximately - but within the intrinsic limit of measurements in
dS space we can't tell the difference) for local physics within a
single horizon.

So we will try to match, as closely as possible, the formulation
of quantum field theory in the global coordinate system on dS
space. QFT, even with some sort of UV cutoff, appears to describe
an infinite number of degrees of freedom in dS space. Any correct
formalism will provide some kind of infrared cutoff, and produce a
finite number of states.

We begin by recalling the thermofield double interpretation of the
Euclidean vacuum state in dS space.   If we consider the generator
of the dS group $H$ corresponding to a fixed boost in
$SO(1,4)$\footnote{Readers who are worrying about the fact that
$SO(1,4)$ has no finite dimensional unitary representations, are
urged to hold on until the next subsection.}, the thermofield
interpretation suggests that we write $H = H_+ - H_-$, where
$H_{\pm}$ are positive definite and represent the Hamiltonians of
the North and South causal diamonds.

However, this does not mean that the generators $H_{\pm}$ must
coincide with the static Hamiltonian of the previous section.  In
fact, we will see that they are more closely related to the Poincare
generators of the local observer.

In particular, since a global coordinate description should make no
reference to the cosmological horizon, our analysis of the
commutation relations between the static and Poincare generators is
no longer applicable.   In global coordinates we should instead
expect that the two groups are related by contraction in almost the
usual way.  The equivocation ``almost" in the previous sentence
refers to the fact that we expect Poincare energies to be bounded
from below.    Thus the correct relation is

$$H_{\pm} \rightarrow R P^0_{\pm}.$$

This then allows us to imagine SUSY generators satisfying
$$[Q^{\pm}_{\alpha}, \bar{Q}^{\pm}_{\dot{\beta}} ]_+ =
\sigma^{\mu}_{ \alpha\dot{\beta}} P_{\mu}^{\pm},$$ for $R
\rightarrow\infty$.

 We do not expect these generators to commute
with the Poincare generators for finite $R$.  Instead the
commutator should go to zero only in the $R\rightarrow\infty$
limit. In order to get some insight into the form of the non-zero
commutator, we consider the super-dS group. The four dimensional
Majorana representation of the Dirac matrices has the following
properties:

$$[\gamma^{\mu} , \gamma^{\nu} ]_+ = \eta^{\mu\nu}$$
$$(\gamma^{\mu})^* = - \gamma^{\mu}$$
$$(\gamma^{\mu})^T = - \gamma^0 \gamma^{\mu} \gamma^0.$$
That is, all matrices are imaginary, $\gamma^0$ is antisymmetric,
and the others are symmetric. $\eta $ is the ``mostly minus" four
dimensional Minkowski metric. $\gamma^4 \equiv i \gamma^0 \gamma^1
\gamma^2 \gamma^3$ is another imaginary anti-symmetric matrix
which completes the algebra of $\gamma^M = (\gamma^{\mu}, \gamma^4
)$ to

$$[\gamma^M , \gamma^N ]_+ = \eta^{MN},$$
the $SO(1,4)$ invariant Clifford-Dirac algebra.   The matrices
$\gamma^0 \gamma^{\mu}$ and $\gamma^0 \gamma^4 \gamma^{MN}$ (where
doubly indexed $\gamma$ matrices are anti-symmetrized products of
singly indexed ones), are all symmetric. The super $SO(1,4)$
algebra is
$$[q_{\alpha}, q_{\beta}]_+ = (\gamma^0 \gamma^4 \gamma^{MN})_{\alpha\beta}J_{MN},$$
where $J_{MN}$ are the $SO(1,4)$ generators.  Although group
theory allows the $q_{\alpha}$ to be real, there is a well known
problem with representing this algebra with Hermitian
$q_{\alpha}$, since there are no highest weight generators of the
dS algebra. Instead, we will write
$$q_{\alpha} = \sqrt{R} (Q_{\alpha}^+ + T^{\alpha\beta}
Q_{\beta}^-),$$ where $T$ represents a time reversal operation and
satisfies $T T^T = - 1$.   The operators $Q_{\alpha}^{\pm}$ are
$SO(1,3)$ Majorana spinors and satisfy the algebra
$$[Q_{\alpha}^+ , Q_{\beta}^- ]_+ = 0$$

$$[Q_{\alpha}^{\pm}, Q_{\beta}^{\pm} ]_+ = (\gamma^0
\gamma^{\mu})_{\alpha\beta} P_{\mu}^{\pm}$$

The anticommutation relations of the large and small $q$
supercharges are consistent in the large $R$ limit if we take
$J_{0\nu} = R(P_{\nu}^+ - P_{\nu}^- )$, and assume that the
$SO(1,3)$ generators are bounded by something of order the
$P_{\mu}$ generators.

Note that the super dS algebra (actually its q deformation) is
represented only in the full thermofield double Hilbert space. It
is amusing that the thermo-field double interpretation of dS space
provides us with a resolution of the old problems of realizing the
dS supergroup in quantum theory. In order to obtain a symmetry
group with finite dimensional unitary representations, I propose
to q-deform the dS super-algebra. We will discuss q-deformation in
the next section, but we note that the operation of q-deformation
does not change the commutators of Cartan generators with raising
and lowering operators.  We will use these undeformed commutators
to estimate the scale of SUSY breaking.

In particular each component of the SUSY generators $Q$ would have
fixed weights, $w$,under commutation with $H^+ - H^-$.

Let us combine this algebra with our expectations for the spectrum
of $P_0$ based on the previous section.   There we argued that the
global QFT picture of dS space might be valid over a region
encompassing of order $R^{1/2}$ horizon volumes as long as we
restricted attention to states well described by QFT.  Those
states had entropy of order $R^{3/2}$ per horizon volume and
energy per horizon volume of order $R$.  The total energy is
extensive, and is of order $R^{3/2}$.   Thus, our Poincare
generators $P_0^{\pm}$ should be  finite dimensional matrices
whose operator norm is of order $R^{3/2}$.  The operator norms of
$Q^{\pm}$ are thus of order $R^{3/4}$.   From the fact that the
weight of components of the SUSY charges is $R$ independent, and
the relation between the dS generators $H$ (in global gauge) and
the Poincare generator $P^0$, it follows that

\eqn{dssus}{[P_0^{\pm}, Q^{\pm} ] =  {w\over R} Q^{\pm}}

Consider a pair of normalized boson and fermion ``particle"
states, localized in a given horizon volume. These are actually
members of a large ensemble of equivalent states.   In static
gauge this is the vacuum ensemble, with one meta-stable
excitation.   In global gauge, we describe it as states consisting
of one excitation in a single horizon volume plus generic field
theoretic excitations in of order $R^{1/2}$ disjoint horizon
volumes.   The typical $P^0$ eigenvalue in this ensemble is of
order $R^{3/2}$.   The masses of single particle excitations must
be extracted by an approximate decomposition: $P^0 \approx \sum
P^0_A$, where the sum is over the $N^{1/2}$ horizon volumes.
Similarly $Q \approx \sum Q_A$.   It is the individual $Q_A$
operators which transform a state with one particle in horizon
volume $A$ into the state with one superpartner in the same
horizon volume, without affecting the states in the disjoint
volumes.   Write the commutator \eqn{susbrk}{[P_0, Q_A ] = {w\over
R}h_A^B Q_B .}  Consistency with the dS supergroup relation
implies $\sum_A h_A^B = 1$.   Locality would suggest $h_A^B \sim
\delta_A^B$, but there is no reason for locality to apply to this
correction to the $R \rightarrow \infty$ limit.   Instead, I
postulate that the matrix elements of $h$ along a fixed row are
all of order $1$, with the same phase. $\sum_A h_A^B = 1$ is
achieved by cancelations within each column of $h$.   Thus $[P_0,
Q_A ] \approx {w\over R} Q$.   The matrix elements of $Q$ in a
typical state of the ensemble are of order $R^{3/4}$, from which
we conclude that the typical splitting between superpartner masses
is of order $R^{-{1\over 4}}$. This is precisely the formula
$\Delta m_{SUSY} \sim \Lambda^{1/8}M_P^{1/2},$ for the maximal
scale of SUSY breaking in dS space, which I conjectured in
\cite{tbfolly}.

The gravitino will be special, because the zero momentum gravitino
state will be related to the action of $Q^+$ on the vacuum. I have
not been able to derive the scaling law for the gravitino mass
from this information alone. Of course, if the model is compatible
with low energy field theory in a single horizon volume (which we
have assumed but not demonstrated in the previous discussion), the
scaling law for the gravitino follows from that for the maximal
SUSY splitting by the usual Ward identity.   In general, one might
expect a class of particles whose masses scaled like that of the
gravitino.

It is worth devoting a few more words to the violation of locality
implicit in our ansatz for $h_A^B$.   Locality in the sense of
commutation of operators at spacelike distances, applies to
systems that can be described by quantum field theory at all
times.    This is not true of our global presentation of dS space.
It is important to realize that the time evolution operator in
global time is {\it not} a member of the dS group.   Furthermore,
our discussion of measurement theory makes clear that at global
time $t=0$, most of the states of the system cannot be described
by field theory.   The entropy in localizable states is only about
$R^{3/2}$.   The rest of the states interact with the localizable
ones with about equal weight.  At a much later time we can
describe all of the $e^{R^2}$ states by quantum field theory, but
we have no more reason to assume that the commutator of the global
$P^0$ with the local $Q_A$ depends only on nearby horizon volumes.
The $Q_A$ are not (simultaneously) defined at $t=0$, and are
evolved from more complicated operators at that time symmetric
point.  At $t = 0$ one can define $Q_A$ for one particular horizon
volume.  The rest of the states are thermal excitations on that
horizon.  The global Poincare generator is not a particularly
transparent operator at that time.

 It is interesting to note that one can,
at the level of precision about the algebraic details that was
used above, generalize this discussion to arbitrary dimension. If
one uses four dimensional formulae to relate the gravitino mass to
the maximal SUSY splitting, one obtains a prediction which does
{\it not} agree with the result of \cite{sushor}.  Of course,
there is a glaring lacuna in this argument.  Low energy, positive
metric SUGRA in dimension higher than $4$ does not admit solutions
with spontaneously broken SUSY and de Sitter vacua\footnote{This
statement is much stronger than the usual no-go theorems which
have been made by string theorists.
 We do not restrict attention to highly supersymmetric low energy Lagrangians, but
 to the most general, minimally SUSic Lagrangian in $d \geq 5$, with positive metric
 excitations. We do exclude solutions of the form $dS_d \times N_p$, with $N$ non-compact.
 If there are quantum theories of such space-times, they have an infinite number of
 states.}.   Thus, there is no consistent low energy way to
 extract a prediction for the gravitino mass from that of SUSY
 matter multiplets in $d \geq 5$ with positive c.c. (which is why we had to
 use a four dimensional relation above).  As I
 emphasized in \cite{tbfolly}\cite{susycosmopheno}, I view
this as an argument that dS space is a strictly four dimensional
phenomenon\footnote{The question of $2$ and $3$ dimensional dS
spacetimes is more confusing.   There are low energy SUGRA
lagrangians with dS solutions.  However, if dS quantum theories have
a finite number of states, they only make unambiguous predictions in
an asymptotic expansion in small c.c.   The predictions with minimal
ambiguity are those for quantities which approach scattering matrix
elements at center of mass energy fixed as the c.c. goes to zero. It
is not clear that there are any super-Poincare invariant
gravitational scattering matrices in less than four dimensions, so
the small c.c. limit is much harder to understand.}.

In summary, the thermofield double interpretation of dS space,
enables us to see how a representation of the dS supergroup could
be compatible with the positivity of energy.  We also saw that
this algebraic structure could lead naturally to the CSB scaling
law for SUSY breaking.   The crucial point here was the assumption
that the commutator between the Poincare Hamiltonian and the
restriction of the SUSY generators to one horizon, was
proportional to the full SUSY generator. This is equivalent to
saying that the breaking of SUSY can be attributed primarily to
states outside (or on) the cosmological horizon of a given
observer.

\subsection{Quantum groups}

It is a trivial mathematical observation that a system with a
finite number of states cannot carry a unitary representation of
the non-compact dS group. In the fall of 1999 A. Rajaraman
suggested to me that a q-deformed version of the dS group might
solve this problem.  B. Zumino confirmed that his student, H.
Steinacker\cite{steinq}, had indeed shown that q-deformed $
SO(2,d)$ groups had finite dimensional unitary representations,
when $q$ is a root of unity. Steinacker's work depended in a
crucial way on highest weight representations, and it was not
clear to me how to generalize it to the dS group. This problem has
recently been solved by Guijosa and Lowe\cite{guylow}. The crucial
observation is that, since only periodic functions of the Cartan
generators appear in the q deformed algebras, we can have finite
dimensional representations which are not highest weight. These
quantum group representations are called {\it cyclic}. Guijosa and
Lowe showed that in $1 + 1$ and $2 + 1$ dimensional dS space,
there is a sequence of cyclic representations of the q deformed dS
algebra, which converges, as $ q\rightarrow 1$ to each principal
series unitary representation of the dS group. The principal
series representations are precisely those which appear in QFT in
dS space.

In order to utilize these ideas to build a quantum model of dS
space, I will use a construction of quantum groups in terms of
creation and annihilation operators, due to
Polychronakos\cite{poly}.   This construction uses a different
definition of the co-product than that widely used in the
mathematics literature, but it appears to be completely consistent
when the quantum group is $U_q (N)$ and the creation operators
transform in the fundamental.   I will want to use it for $SO_q
(1,4)$ with creation operators transforming in a reducible cyclic
finite dimensional unitary representation of this group. $q$ will
be an $N$th root of unity.

We will determine $N$ in terms of $R$ by noting that $SO_q (1,4)$
has an $SO_q (3)$ subgroup which we should think of as rotation of
the dS horizon into itself.   The dS horizon is a holographic
screen and we should think of the finiteness of the dS Hilbert
space as arising from the pixelization of this screen at the
Planck scale. In\cite{susyholoscreen} I argued that quantum pixels
of a holographic screen were fermionic operators, which
transformed in a spinor representation of the transverse rotation
group.  In spherically symmetric situations, they should form a
finite dimensional approximation to the spinor bundle over the
sphere\footnote{That is, each fermion operator should be thought
of as being assigned to a given quantized area on the sphere, and
transform as a spinor under the local tangent space rotations. }If
there are $k$ fermion operators there will be $2^k$ states, so $k$
should scale like the area of the sphere.  To obtain a
construction which realizes this in four space-time dimensions
label the fermion annihilation operators as $N \times N + 1$
matrices,$\Psi_i^A$, with $N\sim R$ in Planck units.  The $SO(3)$
group acts on this in the tensor product of the $[N]$ and $[N +
1]$ dimensional representations. This is the direct sum of all
half integer spin representations up to spin $N + {3\over 2}$.
There is an irreducible action of $SU_q (2)$ in each of these spin
spaces, with $2\pi({\rm ln} q)^{-1} = N + 1$.

A question which I have not resolved is whether or not there is, for
each $N$, a unique and natural unitary representation of $SO_q
(1,4)$ on the space of fermion indices $(i,A)$, which preserves the
anti-commutation relations
$$[\Psi_i^A , (\Psi^{\dagger} )^j_B ]_+ = \delta_i^j \delta^A_B$$
I will assume that there is, and call this representation ${\cal
R}$. Presumably, a way to search for it is to find an appropriate
sub-algebra of the $SL_q (N(N + 1))$ which acts on the fermion
labels. We can then use Polychronakos'\cite{poly} construction to
extend the action of $SO_q (1,4)$ to the entire fermion Fock space.

Now note that since the fermion creation and annihilation
operators transform in the representation ${\cal R}$, we can write
the Fock space as a tensor product of states with positive and
negative weights of $H \equiv J_{04} $.  I will interpret this as
the splitting of the Hilbert space into states of the northern and
southern causal diamonds in dS space.   In (by convention) the
northern diamond, $H$ will have only positive weights.

We now come to a question which is both extremely interesting and
completely confusing.   In the limit in which $N$ (and therefore
$R_{dS}$) go to infinity, does the generator $H$ of the quantum
group approach the static $H$ of the previous section, or $P_0$
the Poincare generator?   On the one hand, we have argued that in
the static patch gauge, the distinction between these two
generators is connected to physics very close to the horizon of a
given observer. In QFT the global gauge physics does not single
out any horizon. In particular, a global observer should, in some
approximation, agree with the QFT prediction that there is nothing
special, like a buildup of the density of localized states, in the
region of the horizon of any given time-like geodesic.

On the other hand, the q-deformation of the group {\it has} picked
out a favored member of the conjugacy class of $H$ by choosing a
particular split into Cartan generators and raising and lowering
operators.   Note that, as far as $H$ is concerned, a similar split
is chosen by the contraction of the dS group to the Poincare group
(We choose $P_0 = {1\over R} J_{04} $ for some particular
non-compact Cartan generator in $SO(1,4)$).

I will take the point of view that the operator $H$, periodic
functions of which appear in the quantum dS group, converges to $
R P_0$ in the $N \rightarrow\infty$ limit.  Similarly the
q-deformed $J_{i0}$ generators should converge to $R P_i$, where
$P_i$ are the Poincare momentum generators.   Thus, with the
association between $N$ and $R$ discussed above, the $q
\rightarrow 1$ limit of the quantum algebra is not the dS algebra
$SO(1,4)$, but its contraction, the Poincare algebra $ISO(1,3)$.

More precisely, we envisage the above limiting procedure taken
only for the generators $H_{\pm}$, where $H = H_+ - H_-$ and the
individual operators are the positive generators in the
thermofield double interpretation of the Hilbert space.  That is,
$H_{\pm}$ acts only on the tensor factor of the Hilbert space
generated by fermion creation operators with positive (negative)
weights, of $J_{04}$. On that factor it coincides with $\pm
J_{04}$.

I believe that there is likely to be a supersymmetric version of
this contraction:  a q-deformed super dS algebra which contracts
to two copies of the Super-Poincare algebra in the $R \sim N
\rightarrow\infty$ limit.  It is easy enough to write down a
q-deformed dS SUSY algebra, but I have not yet found a
construction of $SO_q (1,4)$, or its supersymmetric extension, in
terms of the fermion pixel operators.

 To summarize, I have sketched a formalism for
constructing a finite dimensional theory of global dS space which
approximates the quantum field theory description of this
space-time and is invariant under a quantum deformation of the dS
super-group. The Hilbert space has a natural tensor split into
states with positive and negative values of the static Hamiltonian
of a given causal diamond.  I postulated a similar split for the
fermionic generators and interpreted the split super-generators as
approximations to Poincare super-charges.   Then, given a
hypothetical set of commutation relations for these split charges
in the $R\rightarrow\infty$ limit, I gave a new derivation of the
scaling law for SUSY violation, which I postulated in
\cite{tbfolly}.  To put these results on a firm basis we have to
generalize Polychronakos oscillator construction of $U_q (N)$
quantum groups to fermions transforming in a representation of
$SO_q (1,4)$ and its supersymmetrization.   Then we have to show
that the generators can be split into operators in the positive
and negative weight factors, which satisfy the relations
\ref{susbrk} in the large $R$ limit.

\section{\bf IR divergences in dS/CFT}

{}From the point of view of the semi-classical approximation, the
simplest way to define gauge invariant quantities for quantum
gravity in asymptotically dS space is to define a path integral
with boundary conditions on ${\cal I}_{\pm}$.  This is formally
invariant under coordinate transformations which approach the
identity on the space-like boundaries of space-time.

There are two attempts to define such a formalism in the
literature\cite{witstrom}.   I will use the name dS/CFT, proposed
by Strominger, to refer to both of them.   There has been some
confusion about the relation between these approaches.  The
confusion was cleared up in work of
Maldacena\cite{maldafluct}\footnote{For the most part, Maldacena's
work is not published, though hints of the formalism can be found
in the reference quoted.  I would like to thank J. Maldacena for
explaining his approach to this problem.}.   Maldacena suggests
that the gauge invariant correlation functions are obtained by
first computing the Hartle-Hawking wave function of the universe.
This is the Euclidean path integral of a field theory which
includes gravity, on a manifold with the topology of a $d$
dimensional hemi-sphere. In the semi-classical approximation, the
geometry is that of a hemi-spherical cap, with maximum polar angle
$\theta_M$. The wave functional of the universe is a functional of
the boundary values of fields on this Euclidean manifold.

A baby version of this procedure is the quantization of strings.  By
analogy with that example, one chooses a gauge which is covariant
under the infinitesimal rotation group of the sphere.   In
Maldacena's approach, the dS/CFT correlation functions are defined
by writing

\eqn{hhwf}{{\rm ln} \Psi [\phi] = \sum \int \phi (\Omega_1 ) \ldots
\phi (\Omega_n ) G(\theta_M , \Omega_1 \ldots \Omega_n ).}

$\phi$ is a shorthand for all of the fields in the theory, including
$h_{\mu\nu}$, the gauge fixed fluctuation of the geometry.  Now,
analytically continue $\theta_M \rightarrow {\pi\over 2} + it$ and
take $t \rightarrow\infty$.   If the limiting correlation functions
exist, they will be covariant under the conformal group of the $d-1$
sphere.   The conformally invariant data contained in these limiting
correlation functions (which would be anomalous dimensions and OPE
coefficients, if the correlation functions defined a standard type
of CFT) are the gauge invariant observables of quantum gravity in
asymptotically dS space.

If we believe that quantum gravity in such a space-time has a finite
number of states, this formalism must not be correct outside of
perturbation theory.   Rather, it should result from an illegitimate
approximation to a finite dimensional theory, in the limit of large
Hilbert space dimension.    I propose that a sign of this would be
the appearance of IR divergences in loop calculations of the gauge
invariant data provided by the dS/CFT prescription.   According to
the prescription of \cite{nightmare}, the cosmological constant
itself should provide an IR regulator.   IR divergences should be
reinterpreted as corrections to the classical formulae for the c.c.
dependence of gauge invariant quantities.  In particular, one can
imagine that the formula for the gravitino mass is corrected in the
manner conjectured in \cite{tbfolly}.

The literature on IR divergences in dS space is large\cite{irds},
but it suffers from divergences of opinion, which stem from the
absence of an agreed upon set of gauge invariant quantities.  The
proposal here is that dS/CFT provides us with a definition of such
quantities, and that IR divergences in the dimensions of boundary
fields signal that the classical formulae determining the
cosmological dependence of bulk masses are incorrect.

So far only a warm up calculation has done, which might indicate
that this is so.  Lorenzo Mannelli, Willy Fischler, and
I\cite{tblorir} have done one loop calculations of boundary
correlators in dS space, in theories with massless minimally coupled
scalars, $\phi$.  We found a divergence of the anomalous dimension
of boundary fields dual to massive bulk fields that had soft
couplings to $\phi$.  We are currently trying to sort out the gauge
fixing subtleties of an analogous calculation in gravitational
theories.   The physical components of the graviton propagator have
the same logarithmic behavior as a minimally coupled scalar, but one
must check that the divergences do not cancel in the gauge invariant
boundary dimensions.

It is conceivable that one might get the ``right" formula for the
dependence of the gravitino mass on the c.c. by summing up IR
divergences in dS/CFT.   This would be complementary to the
algebraic approach of the last section.   However, it is not at
all clear that this is possible.  I am fairly confident that
purely field theoretic calculation will show that the field theory
approximation is (contrary to popular opinion) not under control,
but not at all confident that one can use these methods to get the
right answer for quantum gravity.

\section{\bf Phenomenology of CSB}

The earlier sections of this paper outlined my attempts to construct
a quantum theory of a stable dS space and show that the scaling
relation $m_{3/2} \sim \Lambda^{1/4}$ follows from it.   Here, I
want to anticipate the eventual success of those attempts, and
sketch the phenomenology that results.   I have made several
attempts to guess the right phenomenology and will only discuss the
latest, which I believe is the most successful.

There are several general features which must be shared by any
phenomenological model which is derivable from CSB:

\begin{itemize}

\item  1.The model should have,
at the level of the effective Lagrangian, a dS vacuum state, and it
must contain a parameter which allows one to tune the positive
cosmological constant to any value.

\item 2.As $\Lambda \rightarrow 0$ it must become Super-Poincare
invariant.  We call the $\Lambda = 0$ Lagrangian, the {\it limiting
theory}.

\item 3.The dS vacuum state of the low energy field theory model
should be absolutely stable.

\end{itemize}
These conditions can only be realized in four dimensions, with
$N=1$ SUSY, and the limiting theory must be SUSic, and invariant
under a discrete \footnote{since string theory does not have exact
continuous global internal symmetries}  $R$ symmetry which
guarantees the vanishing of the superpotential. In addition, since
we want to be able to turn on SUSY breaking of arbitrarily small
magnitude, by turning on $\Lambda$,   the limiting theory must
have a massless field, which becomes the longitudinal component of
the gravitino when $\Lambda \neq 0$.   Since the SUSY breaking can
be made arbitrarily small, this field must live in a linear
supermultiplet. In the model I will discuss here, it is a chiral
superfield $G$.

To avoid the possibility of decay of the dS minimum to a
non-positive vacuum energy region of field space at large $|G|$, I
will assume that the complex field $G$ lives in a compact space. The
size of this space might be of order $M_P$, or $M_U$, the
unification scale.   This assumption will not be terribly important
for practical purposes, since phenomenology will suggest a much
smaller range of variation of $G$.

I will assume that $G$ has $R$ charge zero, and in addition that it
is charged under a discrete ordinary symmetry ${\cal F}$.   $G^a$ is
the lowest order ${\cal F}$ invariant holomorphic function of $G$.

The effective Lagrangian for the coupling of $G$ to the standard
model has the form

\eqn{Gefflag}{{\cal L} = \int d^4 \theta\ \  G\bar{G}\ K(G/M_1 ,
\bar{G}/M_1, S_{SM}/M_1, \bar{S_{SM}}/M_1 ) }\eqn{Gefflagtoo}{+
\int d^2 \theta\ \sum f_i [(G/ M_1)^a]\ {\rm tr}
(W_{\alpha}^{(i)})^2  + g_{\mu} GH_u H_d + \Lambda^{1/4} M_P^2
f(G/M_P) + c.c.}

Here $S_{SM}$ stands for a generic chiral superfield in the MSSM.
The rest of the standard model is standard. In particular, the
small parameters in the Yukawa couplings of Higgs, quarks and
leptons, are assumed to have been explained by physics (the
Froggatt-Nielsen mechanism?) at the unification scale.  Thus, the
quark and lepton masses and mixing angles are assumed to be
determined, to a good approximation by supersymmetric physics in
the limiting model.

The last term in the Lagrangian is the only one which breaks the
discrete $R$ symmetry.  It arises from interactions with states on
the dS horizon.   We will see that it leads to spontaneous SUSY
breaking.   The scaling of the coefficient of this term has been
determined to agree with the estimate of \cite{sushor} for
$m_{3/2}$. Since the cosmological constant is a high energy input,
one tunes the dimensionless parameters in $f$, in order to set it
equal to its high energy input value.  These two statements are
the primary input of CSB ideas to the effective Lagrangian.

$R$ symmetry charges are chosen\cite{susycosmopheno} so that they
forbid all dimension $4$ and $5$ operators that violate baryon and
lepton number, apart from the dimension $5$ operator that gives
neutrino masses.

The Kahler potential and gauge coupling function, contain irrelevant
terms scaled by the mass $M_1$.   We will assume that this is a
scale generated by as yet undetermined strongly coupled theory,
starting from a weakly coupled effective Lagrangian at the
unification scale. It satisfies $M_1 \ll M_U$. The cross couplings
between $G$ and standard model fields, in the Kahler potential, are
generated by a combination of this new dynamics, as well as the
standard model couplings, and $g_{\mu}$, at scales above $M_1$.

The dynamics at scale $M_1$ arises from a new set of low energy
degrees of freedom, which must be described by some four dimensional
effective field theory below the unification scale. We denote this
theory by ${\cal G}$ .  In order to generate the couplings to the
standard model, which we have included in $K$ and $f_i$, we must
also have relevant or marginal\footnote{possibly marginally
irrelevant} couplings of standard model fields to these new degrees
of freedom. These are assumed to consist of standard model gauge
interactions and a coupling $g_{\cal G} G O_R$, where $O_R$ is a
dimension two operator in the ${\cal G}$ theory at the unification
scale. As an example of such a model, we could take ${\cal G}$ to be
a new SUSic gauge theory, with chiral fields $F_i$ which transform
under both the ${\cal G}$ group and the standard model.   $O_R$
would be a gauge invariant bilinear $c_{ij} F_i F_j$. Unfortunately,
I have not yet been able to find an example of a theory whose
dynamics will generate the effective Lagrangian \ref{Gefflag}.  I
emphasize that the calculation of the Kahler potential and gauge
coupling function can be done, to a good approximation, in the
limiting model.

 The potential for $G$ is
approximately $[V(G/M_1 ,\bar{G}/M_1 ) |f^{\prime} (0)|^2 - 3
|f(0)|^2]\Lambda^{1/2} M_P^2$, where $V =(\partial_G
\partial_{\bar{G}} [G \bar{G}\ K])^{-1}$. A minimum at a value of $|G|\sim M_1$ breaks SUSY,
with a gravitino mass in accord (by design) with the CSB formula.
$|f(0)|$ is fine tuned to guarantee that the low energy
cosmological constant is $\Lambda$ rather than something of order
$\Lambda^{1/2} M_P^2$. From the point of view of CSB this is the
right thing to do. $\Lambda$ is a high energy input even though it
controls the large scale structure of space-time\footnote{Indeed,
{\it precisely because} it controls the large scale structure of
space-time it also controls the spectrum of the highest (Poincare)
energies (black hole states) seen by a local observer.  This is
the UV/IR correspondence. The reader who is confused about which
energy we are talking about should review the previous section.}
and its value cannot be understood from a local effective field
theory, nor modified by renormalization.   Its value in the
effective field theory {\it must} be tuned to agree with the
fundamental definition of the theory.

The couplings of $G$ to standard model fields now give rise to soft
SUSY breaking parameters for the SSM.   Assuming the canonical
estimate $f_i \sim {{\alpha_i}\over \pi}$, we get gaugino masses of
order

\eqn{m1/2}{m_{1/2}^i \sim {{\alpha_i}\over \pi}{\Lambda^{1/4} {M_P
\over M_1}}}

The experimental bound on the wino mass now implies that $M_1 \leq
1 $ TeV which is uncomfortably low.  Pure numerical factors could
be important here.  For example, if we use the reduced Planck mass
in these estimates, the bound on $M_1$ is reduced by a factor of
$\sqrt{8\pi}$, and the model is definitely in contradiction with
experiment. On the other hand, we do not have a sharp estimate for
either $f_i$ or the horizon induced superpotential, which
determines $F_G$. Either of these could raise the estimate of
$M_1$.

I should emphasize that, as an exercise in pure math, we can imagine
doing the calculation of the effective Lagrangian for very small
values of $\Lambda$.   The estimate of $M_1$ we just made, suggests
that this approximation may be inappropriate (or borderline) for the
real world value of $\Lambda$.   That is, in order to make
calculations appropriate to the real world we may need to treat more
details of the dynamics of the ${\cal G}$ theory than can be
incorporated in an effective Lagrangian for the MSSM coupled to $G$.

Terms in $K$ of the form $G\bar{G} |S|^2 $ where $S$ is some SSM
chiral superfield, will contribute to squark, slepton and Higgs
boson mass terms when we insert $F_G$.   The contributions to squark
and slepton squared masses will be of order $({\alpha_i \over
\pi})^2 {|F_G|^2 \over M_1^2} $ and come from standard model gauge
loops above the scale $M_1$, as in gauge mediation.  They will be
flavor diagonal and positive. The Higgs masses will, in addition,
get contributions which depend on the Yukawa couplings $g_{\mu}$ and
$g_{\cal G}$.  If these are relatively strong they will lead to a
negative contribution to the squared Higgs mass. If the $\mu$ term
is not too large, this leads to spontaneous breaking of electroweak
symmetry. The $\mu$ term in the SSM arises through the VEV of $G$
and has the right order of magnitude if $|G| \sim M_1 \sim $ TeV .

SUSY CP violation is not problematic in this model.  Indeed, if
the ${\cal G}$ theory has {\it automatic CP conservation} then the
model even solves the strong CP problem of
QCD\cite{susycosmopheno}. Automatic CP conservation in a gauge
theory means that one can rotate away all of the CP violating
phases, including the ${\cal G}$ gauge theory vacuum angle.  In
the limit $\Lambda \rightarrow 0$, the supersymmetric low energy
theory has no QCD vacuum angle because of the anomalous gluino
chiral symmetry.  When $\Lambda \neq 0$, and if the ${\cal G}$
theory is automatically CP conserving, then, apart from the CKM
matrix, CP violation appears only through a phase in $F_G$.  This
infects both the gluino mass and the B term, $F_G h_u h_d$, in a
correlated way.  However, using the fact that the number of
generations equals the number of colors, we can rotate away both
of these phases without introducing $\theta_{QCD}$.  The chiral
symmetry which rotates these phases away has no QCD anomaly.

The model has no cosmological moduli problem, because the $\Lambda
\rightarrow 0$ limit must be an isolated $N=1$ vacuum, apart from
the compact $G$ field.  Similarly it does not have a cosmological
gravitino problem.   However, because the gravitino is very light,
and relatively strongly coupled, there is no SUSY dark matter.
Axion dark matter is also ruled out because the freezing of moduli
leaves us with no axion candidate\footnote{Actually, the real part
of $G$ could be an axion candidate, but it would be ruled out by
beam dump experiments.}. The simplest possibility for dark matter
is to postulate some sort of cosmologically stable ${\cal G}$
hadron $B_{\cal G}$ which is a singlet of the standard model gauge
group. The stability of $B_{\cal G}$ could be due to an accidental
symmetry like the baryon number symmetry of QCD.   The mass of
this particle would probably be in the multiple TeV range and its
annihilation cross section would similarly be of order ${1\over (x
{\rm TeV})^2} $, with $x$ a number of order one.   This gives a
freeze out density within shouting distance of the correct dark
matter density, but the details are obviously important.  Also,
since the stability of this particle is due to an approximate
symmetry, its relic abundance will depend on the asymmetry in its
approximately conserved quantum number that is generated in the
early universe.  It is probably necessary to assume that physics
at or above the TeV scale generates an asymmetry in $B_{\cal G}$
in order to get the right relic density of these dark matter
candidates

To summarize, this low energy effective description gives a model of
TeV scale particle physics which is in rough accord with all data,
and avoids many of the problems of conventional SUSic models.   It
has two principal theoretical defects:

\begin{itemize}

\item The scales $M_1$ and $\Lambda^{1/8} M_P^{1/2} $ (the maximal splitting in
SUSY multiplets) are coincidentally close to each other.   One
should remember however that we view $\Lambda$ as a variable input
to the theory, whereas $M_1$ is a parameter which is
calculable\footnote{I am assuming here that the limiting theory is
unique or that there are a very small number of possibilities.  See
the next section for more discussion of this point.}, and has a
finite $\Lambda \rightarrow 0$ limit.  Thus, it does not vary very
much as $\Lambda \rightarrow 0$.   Assume further that the amplitude
of primordial fluctuations at horizon crossing, $Q$ is similarly
independent of $\Lambda$.   Then Weinberg's galaxy formation
bound\cite{wein} bounds $\Lambda $ from above by a number that is
determined by $M_1$ (the dark matter density at the beginning of the
matter dominated era). As $\Lambda \rightarrow 0$ supermultiplets
become degenerate, and the gravitino becomes lighter and more
strongly coupled to the standard model.  At a certain point, nuclei
become unstable to decay into a Bose condensate of snucleons.  This
puts a lower bound on $\Lambda$ of order $10^{-22}$ times its
``actual" value.  An even tighter lower bound comes from requiring
that electroweak symmetry breaking occur.  I am not sure what the
tightest lower bound on $\Lambda$ is.   Indeed, the definition of
``tightest" depends on what facts about the real world one is
willing to put in as input. In CSB  $\Lambda$ is an input parameter
to be constrained by data. Since it is inversely proportional to the
logarithm of the number of states in dS quantum gravity, {\it a
priori} estimates of the probability distribution for small
$\Lambda$ will depend on a ``metaphysical" model, which describes
many possible universes, only one of which will ever be observed (by
us).  One such model was described at the end of \cite{holocosm4}.
It favors large values of $\Lambda$.   If the {\it a priori}
distribution for $\Lambda$ favors larger values, then the existence
of galaxies predicts the coincidence between $M_1$ and the largest
scale of cosmological SUSY breaking.   I emphasize that this
conclusion depends on the extra assumption that the $\Lambda
\rightarrow 0$ model is unique\footnote{Or that there is a small
class of possibilities, one member of which describes the universe
we live in, while the other members do not have galaxies (perhaps
because they have no cosmologically stable massive particles).}.

\item The most serious defect of our model is that I have not yet
found a candidate for the ${\cal G}$ theory.   That field theory
must generate the dynamical scale $M_1$ without giving $G$ a mass
of order $M_1$. It must preserve SUSY and the discrete symmetries,
$R$ and ${\cal F}$. It must not lead to a superpotential for $G$,
and it must not predict extra massless degrees of freedom which
are inconsistent with experiment. Discovering the ${\cal G}$
theory is the most serious unfinished task in phenomenological
CSB.   Its dynamics determines many of the detailed predictions of
our low energy effective Lagrangian.

\end{itemize}

\section{Isolated Poincare invariant models of quantum gravity with four
supercharges}

A key ingredient in our arguments is the notion of the limiting
model, the super-Poincare invariant S-matrix which arises in the
limit $\Lambda \rightarrow 0$.   I have argued that this has to be
an isolated model, meaning that it has a compact moduli space. We do
not have any examples of a model of this type coming from string
theory.

The reason for this is not hard to find.   In $N = 1$ SUGRA,
supersymmetry is generic, but super-Poincare invariance is not. The
equations for a supersymmetric model are $D_i W = 0$.  This is the
same number of equations as unknowns.   The addition of D- terms
does not change this result.   Super-Poincare invariance requires $W
= 0$ in addition.

There are two general strategies for finding Super-Poincare
invariant vacua.  The first, due to Witten\cite{wittenf} relies
explicitly on a non-compact moduli space of approximate vacua. Let
$S$ be the non-compact direction.   Assume further that the
asymptotic range of $S$ is a strip with periodic boundary
conditions, like the moduli space of complex structures of a two
torus.   Then the superpotential has the form $W =
\sum_{n=0}^{\infty} a_n e^{in S}$.   Often, this is precisely the
form of a BPS instanton sum: {\it e.g.} the imaginary part of $S$ is
the volume of some cycle, and the instanton is a brane wrapped on
this cycle.   If one can do the instanton calculation, and show that
$a_n = 0$ for all $n$, then one has established the existence of a
moduli space of $N=1$ vacua.

The second strategy, first studied in \cite{bdmoduli} invokes an R
symmetry, which may be discrete.   In units where the superspace
coordinates have charge $1$, we require that there be no fields of R
charge $2$, and at least one field (again call it $S$) of R charge
0.  Then the range of $S$ is an $N = 1$ moduli space.   In this
case, there is no apparent barrier to assuming that the moduli space
is compact.   Perturbative string theory or low energy SUGRA
approximations cannot find such points.   Those methods rely on an
approximate non-compact moduli space, and an expansion around
infinity in that moduli space.   If, using these approximations, we
find a vanishing superpotential on some submanifold of the moduli
space, to all orders in that expansion, (and an argument that $W=0$
non-perturbatively) then we have constructed a non-compact moduli
space of $N=1$ theories, rather than an isolated one\footnote{Recall
that the the reason for insisting on a compact moduli space, was in
order to ensure that the model with non-vanishing $\Lambda$ cannot
decay, since decay to a zero c.c. states implies an infinite number
of states. Thus, I am simply recapitulating the argument that any dS
state constructed in perturbative string theory will be
meta-stable.}.

Dine\cite{ds} has long argued that discrete moduli spaces should
exist as points of enhanced duality symmetry.  For purposes of CSB
(at least with the low energy Lagrangian we have studied in this
section) we actually need a smooth compact moduli space, the target
space of the Goldstino field, $G$.   It seems clear that if such
moduli spaces exist, they are few and far between.   This makes it
difficult to find them, but suggests that the predictions they lead
to will be pretty unique.

There are two possible approaches to finding compact moduli spaces.
The equations $W = D_i W = 0$ for a section of a line bundle over a
Kahler manifold, are topological, in the sense that they are
independent of the metric on the manifold.   Perhaps there is some
generalization of topological string theory (or rather a
non-perturbative completion of it) which finds solutions of these
equations for compact moduli spaces of super-Poincare invariant
models of 4 dimensional quantum gravity.

A more promising approach is to find non-perturbative dynamical
equations for super-Poincare invariant models of quantum gravity. At
the moment, our only examples of this come from Matrix-Theory and
this only works for a subclass of models, with at least 16
supercharges\footnote{It also requires us to take a difficult large
$N$ limit to achieve super-Poincare invariance.}.   A more covariant
arena for such models is a formulation on null infinity.   Such a
formulation cannot be a standard Hamiltonian quantum mechanics, with
one of the coordinates of null infinity playing the role of time. We
have to find an alternative set of dynamical equations for the
S-matrix\footnote{In four dimensions, we also have to find a more
general object, along the lines of old work by Fadeev and Kulish, to
deal with the problem of infrared divergences in the gravitational
S-matrix.}.

Many years ago, S-matrix theorists hoped that the principles of
analyticity, crossing symmetry, Lorentz invariance and unitarity,
might completely specify the S-matrix.   Mechanically the idea is
(thinking perturbatively) that the unitarity equation for the
T-matrix

\eqn{unit}{i (T - T^{\dagger}) = TT^{\dagger}} would allow one to
calculate the discontinuities across cuts of higher order
amplitudes in terms of lower order amplitudes.  Then generalized
Cauchy Theorems (dispersion relations) would allow one to
calculate the full amplitude, and proceed by induction.   If this
were really true, one might suspect that the non-perturbative
solution of these equations was unique.   We know of course that
this program was doomed to failure, because of the existence of
many consistent quantum field theories.   We also know that in a
standard perturbative approach, even the additional constraint of
local supersymmetry does not fix the S-matrix uniquely.   Order by
order in the low energy expansion of supergravity, expanded about
a supersymmetric Minkowski vacuum one can find consistent
solutions of these constraints with an ever growing number of
arbitrary constants.

The analyticity postulate of S-matrix theory was never given a clear
mathematical formulation, and the rules explicitly excluded massless
particles and assumed high energy behavior of amplitudes which is
not consistent with what we know about gravitation.   Perhaps we
need to revisit these questions, in order to find the proper
formulation for models with compact moduli space.

Indeed, the formulation of non-perturbative rules for determining
the S-matrix of super-Poincare invariant space-time, is an
outstanding problem of String Theory.  We have perturbative rules
in various regions of moduli space, and the
Matrix-theory\cite{bfss} formulation for certain regions of moduli
space.  The latter is non-perturbative, but not fully satisfactory
because it does not manifest super-Poincare invariance before
taking the (difficult) large $N$ limit.   One might imagine that
some of the constraints on super-Poincare invariant S matrices
would be seen in a formulation like Matrix theory, only as a
failure of certain models to have a Lorentz invariant large $N$
limit.  One would also like to shed light on the question of
whether there are Poincare invariant theories of quantum gravity
which are not Super-Poincare invariant.

\section{\bf The relation to string theory}

String theorists constantly ask what the relation of all of these
ideas is to ``String Theory".  In my opinion, the hidden
assumption behind this question is the idea that string theory is
one Hamiltonian with many superselection sectors, and I think this
idea is misguided.   Nonetheless, we could try to give some
meaning to the question.

I think one thing that cannot be true is that a stable de Sitter
space has unambiguous ``gauge invariant" observables, like any of
the conventional string theories we know.   This follows if one
believes the claim that the system has only a finite number of
states, and I have discussed this point extensively above. Instead
I've suggested an equivalence class of theories of de Sitter space
where some of the observables are universal in the large $R$
limit. These are the matrix elements of an ``S-matrix" which
becomes the S matrix of a super-Poincare invariant limiting theory
as $R \rightarrow \infty$.  The ambiguities in appropriately
constrained matrix elements at finite $R$, are of order $e^{ -
R^{3/2}}$.  The phrase {\it appropriately constrained} in the
previous sentence means: a process in asymptotically flat space,
whose S-matrix is well approximated by finite time amplitudes in a
region whose linear size is $\ll R$.

I would like to believe that the limiting model shares properties
of the most realistic string compactifications we know, the {\it
M-theory on G2/Heterotic on CY3/F-theory on CY4} moduli space. and
that it is relatively unique\footnote{Suffiently unique that one
can distinguish the correct model of our world, from the class of
all mathematically consistent theories, by relatively simple
criteria like the existence of galaxies.}.

There is a hypothetical way in which string theory could lead us
to a model of dS space by calculation instead of analogy.   The
basic idea is to construct some sort of brane in asymptotically
flat space, whose local geometry is de Sitter, and find a
decoupling limit where a finite number of states associated with
the brane, cease to interact with the rest of the system.   The
decoupling cannot be precise, because there is a gap in the energy
spectrum of the dS Hamiltonian, so a low energy limit cannot be
taken.  However, since the gap is exponentially small for large
radius, it might be possible to have approximate decoupling. The
residual, exponentially small, couplings to the rest of string
theory, would be examples of the imprecision in the definition of
the dS Hamiltonian.  Different states of the external string
theory, would induce slightly different Hamiltonians on the brane.
In a situation like this, it is likely that the dS brane will be
unstable, but if its lifetime is longer than the tunneling times
of field theoretic apparatus in dS space ($e^{R^{3/2}}$) then it
would enable us to extract a dS Hamiltonian with the degree of
precision measurable in dS space. It may be that the results of
\cite{eva} should be viewed as attempts to implement this
decoupling strategy.

\section{Conclusions}

Cosmological SUSY Breaking is a work in progress.   While I believe
that I have uncovered tantalizing hints of an elegant formulation of
it, they are incomplete.  The first new result in the present paper
was a clarification of the relation between Poincare and static
Hamiltonians in the static gauge.  This led to a successful
prediction of the relation between black hole mass and entropy for
dS black holes. I review the construction\footnote{in collaboration
with B. Fiol,}\cite{davis} of a quantum model of the static gauge
Poincare generator, which reproduced the properties of Schwarzchild
de Sitter black holes. I then proposed a new algebraic program for
constructing a theory of stable dS space in global gauge, and
outlined a possible algebraic derivation of the CSB scaling
relation.   A final new result was an explanation of the coincidence
between the dynamical scale $M_1$ and the CSB scale $\Lambda^{1/8}
M_P^{1/2}$, in the low energy phenomenology that follows from CSB.

 There are, as I see it, three or four
lines of attack on the proposal of CSB:

\begin{itemize}

\item A direct assault on the description of dS in global gauge.
Here one must construct the q-deformed super dS generators and
verify that they can be decomposed into approximate Poincare
super-generators for North and South causal diamonds.  The crucial
results will be to show that the bound on the spectrum of the
Poincare Hamiltonian, $P^0$, scales like $N^{3/2}$ and that the
leading term in the commutator between $P^0$ and the restriction
of the SUSY generator to one horizon, scales like the full SUSY
generator.

\item  Continued investigation of the low energy phenomenology
suggested by CSB.  The crucial step here is to find an appropriate
strongly coupled theory, ${\cal G}$.

\item Construction of a Non-Perturbative algorithm for
Super-Poincare invariant S-matrices in quantum gravity, in order to
investigate the existence of isolated models in four dimensions with
minimal SUSY.

\item An attempt to derive the CSB scaling relation by summing up
infrared divergences in perturbative dS/CFT.   I am somewhat
skeptical that this can give the correct result. The dS/CFT
calculations may demonstrate the existence of large corrections to
the naive classical relation between the gravitino mass and the
c.c., but they may not be sufficient to extract the true behavior in
quantum gravity.

\end{itemize}

\section{Acknowledgments}

I benefitted from conversations with W.Fischler, Mark Srednicki,
Lenny Susskind, and R. Bousso.  I'd particularly like to thank B.
Fiol for collaboration on the fermionic model of the dS black hole
spectrum.

The research of T.B was supported in part by DOE grant number
DE-FG03-92ER40689.

%


\end{document}